%% file: bare_jrnl_new_sample4.tex
\documentclass[lettersize,journal]{IEEEtran}
\usepackage{amsmath,amsfonts}
\usepackage{algorithmic}
\usepackage{algorithm}
\usepackage{array}
\usepackage[caption=false,font=normalsize,labelfont=sf,textfont=sf]{subfig}
\usepackage{textcomp}
\usepackage{stfloats}
\usepackage{url}
\usepackage{verbatim}
\usepackage{graphicx}
\usepackage{cite}
\usepackage{pgfplots}
\pgfplotsset{compat=1.18}
\usepgfplotslibrary{groupplots}
\usetikzlibrary{patterns}
\usepackage{tikz}
\usepackage{caption}
\usepackage{subcaption}
\usepackage[most]{tcolorbox}

\definecolor{lightgraybox}{RGB}{245,245,245}

\newtcolorbox{takeawaybox}{
    colback=lightgraybox,
    colframe=gray!60,
    boxrule=0.5pt,
    arc=2mm,
    left=2mm,
    right=2mm,
    top=1mm,
    bottom=1mm
}

\newcommand{\Description}[1]{}

\begin{document}

%\title{From Federated LLM Fine-Tuning %Traffic to Optical Circuit Switching in 5G %and Beyond RANs}

%\title{Federated LLM Fine-Tuning and Optical Circuit Switching: \\ Can We Get the Best of Both Worlds ?}

%\title{Federated LLM Traffic and Optical Circuit Switching: \\Can We Get the Best of Both Worlds?}

\title{Federated Learning for LLMs over Mobile Networks: \\Issues and Solutions in the RAN Transport}
\author{
    \IEEEauthorblockN{Emilio Paolini$^{1}$ \quad Andrea Pinto$^{2}$ \quad Flavio Esposito$^{2}$ \quad  Luca Valcarenghi$^{1}$ }
    
    \IEEEauthorblockA{$^1$\textit{Tecip Institute, Scuola Superiore Sant'Anna, Pisa, Italy}}
    
    \IEEEauthorblockA{$^2$\textit{Computer Science Department, Saint Louis University, St. Louis, USA}}

    %\IEEEauthorblockA{emilio.paolini@santannapisa.it}

\thanks{This work has been sponsored by NSF Awards \#2133407 and \#2201536, and in part by the CHIPS-JU SMARTY project (GA101140087).}
}


\maketitle

\begin{abstract}
Federated LLM fine-tuning enables large models to be adapted using private and geographically distributed data at the network edge, creating recurring and deadline-sensitive communication workloads across access and transport networks. This challenge is particularly relevant in mobile RANs, where wireless variability, mobility, and device heterogeneity cause model updates to arrive asynchronously. Although these updates belong to the same learning round and share a common destination and deadline, conventional transport networks treat them as independent device-originated flows, hiding their underlying structure and limiting the ability to efficiently provision transport resources. This mismatch is particularly problematic for optical circuit switching and all-photonics transport, which benefit from predictable and schedulable traffic demands. We argue that future RANs should act as learning-aware traffic shapers by exposing the communication structure of distributed model adaptation to the transport layer. Through in-network aggregation at the gNB, asynchronous UE updates can be transformed into fewer aggregate transfers with bounded size and delivery requirements. Once shaped in this way, federated LLM traffic becomes a suitable candidate for selectively provisioned optical connectivity, where high-capacity paths can be established during aggregate-transfer windows and released between learning rounds. The resulting architecture combines the flexibility of packet-based mobile access with dynamically provisioned optical capacity, illustrating a broader approach for coordinating distributed AI workloads across programmable access and transport networks.
\end{abstract}


\section{Introduction}
Foundation models, including Large Language Models (LLMs), are
increasingly moving toward distributed adaptation over private data
generated at the network edge~\cite{cai2023efficient}. Federated
Learning (FL) supports this shift by allowing clients to exchange model
updates rather than raw data~\cite{mcmahan2017communication}, while
parameter-efficient techniques such as Low-Rank Adaptation (LoRA)
reduce the number of trainable parameters communicated during
adaptation~\cite{kuo2026flasc}. Nevertheless, federated LLM adaptation
still produces recurring and deadline-sensitive transfers whose value
depends on whether updates arrive in time to contribute to the current
learning round. This challenge is particularly relevant in mobile Radio Access
Networks (RANs), where wireless variability, mobility, and device
heterogeneity cause updates to become ready asynchronously
\cite{amiri2020federated}. Conventional RAN transport does not expose
their learning-level relationships: each user equipment (UE) forwards
its update independently, while the transport network observes ordinary
device-originated flows. This hides the fact that these transfers share
a learning process, destination, and delivery deadline. Such traffic is also poorly matched to optical-native transport.
Optical Circuit Switching (OCS) and All-Photonics Network (APN)
architectures are most effective when demands are sufficiently
predictable and schedulable
\cite{saliou2025optical,nakagawa2024multi,avin2025revolutionizing}.
UE-level updates instead arrive asynchronously and may be individually
too small or dispersed to justify optical reconfiguration. Additional
capacity alone is therefore insufficient: the RAN must first reshape
learning traffic into demands that the transport network can understand
and schedule.

This motivates a learning-aware role for the RAN. In-Network
Aggregation (INA), originally explored in datacenter networks to
combine model-update fragments in transit~\cite{sapio2021scaling}, can
be applied at the gNB to aggregate updates belonging to the same
learning process before forwarding them upstream
\cite{paolini2026programmable}. The transport network then observes
fewer gNB-to-parameter-server aggregates rather than many asynchronous
UE-originated transfers. The impact extends beyond bandwidth reduction. Each aggregate has a
bounded volume, known endpoints, a readiness time, and a latest useful
delivery time. INA therefore converts device-level variability into
schedulable transport demands and creates a control point for deciding
whether an aggregate should be sent immediately over the packet path,
coordinated with other aggregates, or carried over a temporarily
provisioned optical circuit. This reshaping is particularly attractive because distributed AI
training alternates between computation and synchronization phases
\cite{rajasekaran2024mltcp}. By absorbing part of the UE-level
asynchrony, the gNB can create temporally localized aggregate transfers
that are better suited to selectively provisioned optical connectivity,
with quieter intervals emerging between learning rounds.

Building on this observation, this article investigates how
learning-aware traffic shaping at the gNB can bridge federated LLM
adaptation and optical-native transport. We evaluate how gNB-based
aggregation transforms asynchronous UE uploads into bounded
gNB-to-parameter-server transfers and quantify its effects on learning
progress, update delivery, backhaul traffic, and the temporal structure
of the resulting workload. We further discuss a coordination framework
in which the RAN exposes each aggregate through its volume, readiness
time, source, destination, and latest useful delivery time. A Window
Scheduler coordinates ready aggregates, while an APN Controller
determines whether their combined traffic volume and deadline slack
justify optical provisioning. Although federated LLM fine-tuning is the motivating workload of this
article, the proposed transport abstraction is not model-specific. It
applies more broadly to federated-learning workloads whose updates can
be aggregated at the network edge and exposed as bounded,
deadline-constrained transfers. LLM adaptation is particularly relevant
because PEFT substantially reduces the computation and communication
requirements of full-model fine-tuning, while still generating recurring
transfers that may range from a few megabytes to hundreds of megabytes.
Such traffic volumes can materially affect packet-versus-OCS
provisioning and therefore benefit from coordinated access and optical
transport.

This analysis shows that INA should not be interpreted only as a
mechanism for reducing transmitted bytes. Its broader contribution is
to expose the structure of distributed-learning traffic and enable
deadline-aware coordination across radio, packet, and optical domains.
The resulting transport decision is workload-dependent: large
aggregates may justify an optical circuit as soon as they become ready,
whereas smaller aggregates may require coordination across multiple
gNBs to amortize the optical setup cost. Such coordination must also
account for uncertain readiness times, heterogeneous wireless
conditions, competing learning jobs, fairness, and the risk that
waiting for additional traffic may cause ready updates to miss their
delivery deadlines. Future programmable access networks can therefore better support
distributed AI by transforming application-level communication into
bounded, semantically meaningful, and schedulable aggregate transfers.
The RAN becomes not only the point at which devices access the network,
but also the point at which learning demands are shaped for coordinated
use of the end-to-end communication infrastructure.

\begin{figure*}[t]
    \centering
    \includegraphics[width=0.9\linewidth]{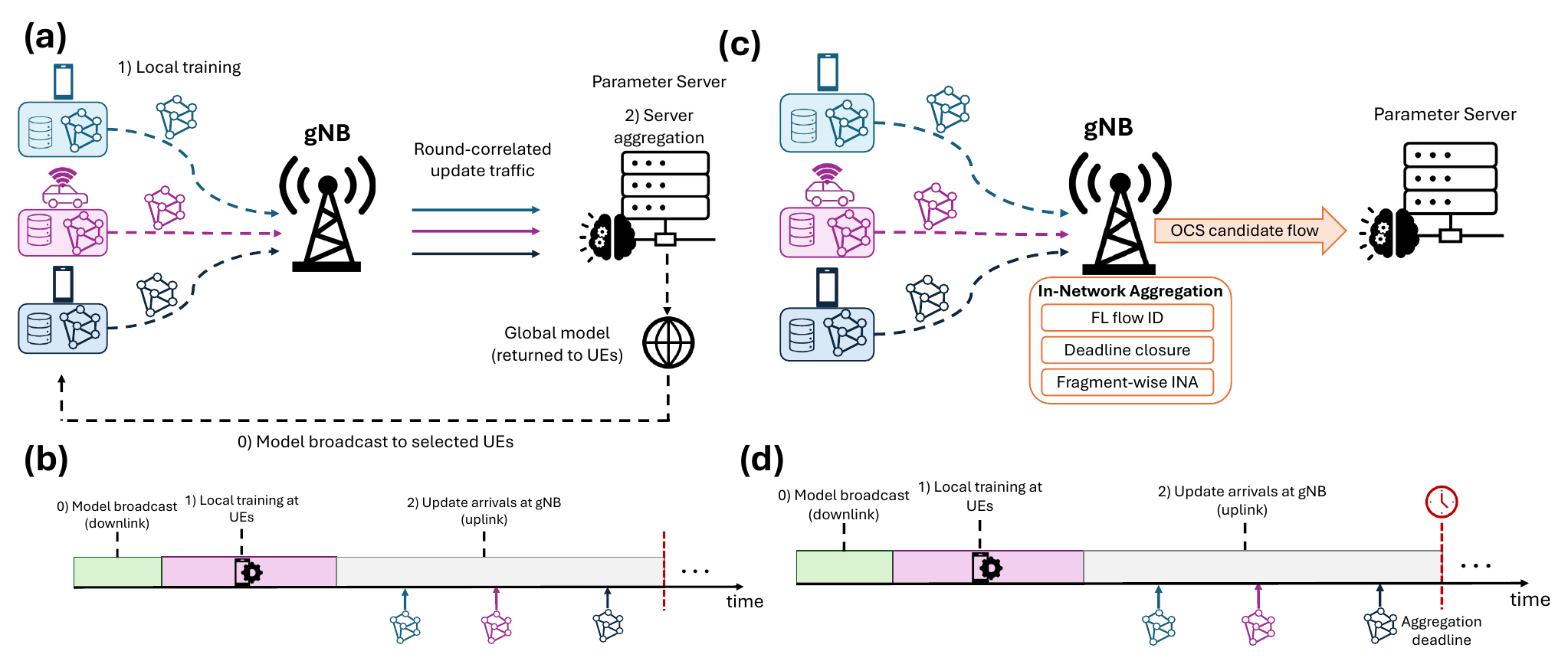}
    \caption{(Left) Federated LLM fine-tuning over the RAN: (a) learning workflow and (b) traffic timeline. (Right) Learning-aware gNB aggregation: (c) processing pipeline and (d) transport timeline.}
    \label{fig:fl-ran-process}
\end{figure*}

\section{Impact of Learning Traffic on the RAN}

While privacy attacks have been demonstrated on FL systems, it is understood that FL preserves some level of privacy during training, as devices keep data locally, compute model updates, and send those updates to a parameter server~\cite{FSL-big-data-Flavio}. When FL clients are connected through a mobile network, the same process acquires an additional meaning: model adaptation generates a recurring transport workload.
Figure~\ref{fig:fl-ran-process}(a) illustrates such learning workflow. Selected UEs receive the global model, train locally, and upload their updates through the gNB to the parameter server. Figure~\ref{fig:fl-ran-process}(b) shows the corresponding traffic timeline: model distribution is followed by local computation and by staggered update arrivals constrained by a common aggregation deadline. The server does not simply receive unrelated application flows; it waits for enough useful updates to construct the next global model.

\begin{takeawaybox}
\textbf{The Main Insight: Structured Learning Traffic.}
A programmable RAN should treat FL updates as a distributed computation rather than as independent network flows. Updates generated during the same training round share a common destination, a common aggregation deadline, and a common semantic objective, constructing the next global model. Exploiting this hidden structure enables the network to jointly optimize communication efficiency and learning performance.
\end{takeawaybox}

Unlike conventional mobile traffic, where each flow can be managed independently, federated learning traffic consists of multiple correlated updates that jointly contribute to the same training round. Consequently, the value of a model update depends not only on successful delivery, but also on timely delivery before the aggregation deadline. An update that arrives too late may still reach the server, yet it can no longer influence the current model and must wait for a future round or be discarded.

The resulting communication workload exhibits three key characteristics.
First, it is \emph{round-correlated}: many updates are generated during the same training round and converge toward a common aggregation point.
Second, it is \emph{deadline-sensitive}: the parameter server must eventually close the round, even if some devices have not completed their uploads.
Third, it is \emph{heterogeneous}: UEs differ in radio conditions, computational capabilities, mobility, energy availability, and local data distributions~\cite{kairouz2021advances}. As a result, the network influences not only communication latency but also which clients ultimately contribute to the global model.
These characteristics create a trade-off between learning quality and communication efficiency. Waiting longer allows more devices to participate, improving data diversity, while closing the round earlier reduces latency but may exclude slower clients and bias the model toward users with better connectivity. Consequently, communication performance depends not only on traffic volume but also on the timing and coordination of model updates.

Today's RAN, however, is largely unaware of this structure. The gNB schedules and forwards packets as independent flows, without considering that many updates belong to the same learning round and share a common objective. As a result, the network currently misses the opportunity to exploit information already present in the federated learning process.

\section{The gNB as a Learning-Aware Aggregation}

The previous section showed the limitation of the conventional transport view: the gNB is the first point where learning updates converge, but it normally forwards them as separate upstream flows. 
The paradigm shift is to use this convergence point as part of the communication phase of training. 
Instead of forwarding one complete update per UE toward a remote parameter server, the gNB can combine updates that belong to the same learning round and forward one aggregate update upstream. The clients still train locally, and the parameter server still coordinates the global model. 
What changes is the role of the RAN. 
The gNB is no longer only a radio scheduler and forwarding point; it becomes a learning-aware aggregation point. It uses the round structure of FL to organize update traffic before it reaches the transport network.
\begin{takeawaybox}
\textbf{Exploit the structure of FL traffic at the gNB.}
The proposed change is to move aggregation from the parameter server toward the network edge. The gNB can use the shared round structure of FL to convert many UE-level updates into a small number of predictable, deadline-aware transfers before they enter the transport network.
\end{takeawaybox}
Figure~\ref{fig:fl-ran-process}(c) shows such a change in the learning workflow. 
On the UE-facing side, update arrivals remain variable because devices finish local training at different times and experience different radio conditions. 
However, these arrivals are no longer exposed one by one to the upstream transport network.

They are absorbed at the gNB and combined into an aggregate that represents the contribution of that gNB for the current round. The box inside Fig.~\ref{fig:fl-ran-process}(c) shows the internal gNB pipeline that makes this transformation possible. First, \emph{FL flow identification} separates model-update traffic from ordinary mobile traffic, for example through a dedicated QoS flow or a user-plane classification rule. Second, \emph{deadline closure} determines the conditions under which the aggregate must be released: the gNB closes it when enough useful contributions have arrived or when the round deadline is reached, rather than waiting indefinitely for the slowest UE. Third, \emph{fragment-wise In-Network Aggregation (INA)} combines update fragments as they arrive, avoiding the need to buffer a complete update from every UE before aggregation.
The gNB cannot wait indefinitely for the slowest UE; it closes the aggregate when enough useful contributions have arrived or when the round deadline is reached.

Figure~\ref{fig:fl-ran-process}(d) shows the corresponding transport timeline. 
Before aggregation, the gNB-to-parameter-server link sees many UE-originated transfers spread over the update-arrival interval. 
With INA, the gNB first collects and combines updates during an aggregation window and then releases one aggregate transfer upstream. 
The transport-facing demand therefore changes from many device-level flows to a smaller number of gNB-level aggregate transfers. This does not require the wireless edge to become perfectly synchronized. 
Some UEs may still be late, and some updates may miss the current round. 
The role of INA is not to eliminate device-level variability, but to contain part of it at the gNB. 
The gNB becomes the boundary where variable UE-level arrivals are converted into structured gNB-to-server demand. 

This abstraction shift is important for optical transport. 
The output of the gNB is not just a smaller amount of traffic; it is a better-defined transport request. The aggregate consolidates multiple UE-originated flows into a single gNB-to-parameter-server transfer with an estimable size and a delivery deadline inherited from the learning round. These are precisely the properties that an APN/OCS controller needs to decide for performing the optical transport network reconfiguration. The FLAG architecture is one example of this principle~\cite{paolini2026programmable}. FLAG embeds aggregation inside the 5G gNB user plane, allowing model updates from multiple UEs to be combined before being forwarded upstream. The goal is not to replace FL, but to make the network aware of its communication structure. Section~\ref{sec:initial-results} quantifies the resulting learning-level
benefits, including the reductions in communication time and overall round duration.

\section{Optical-Native RANs: A New Transport Opportunity}

Optical transport can exploit learning traffic when four pieces of information are known: source, destination, traffic volume, and transmission window. Raw UE-level FL traffic does not naturally provide this description because updates are distributed across many devices and affected by wireless variability. A gNB aggregate is different: its source is the gNB, its destination is the parameter server, its size can be estimated once the aggregation group is known, and its deadline is inherited from the learning round. This does not mean every update should use an optical circuit. Packet switching remains appropriate for control traffic, model distribution, small updates, and fallback operation. Optical provisioning becomes attractive only when the aggregate is sufficiently large and the expected transmission gain outweighs the circuit reconfiguration cost. INA is therefore not a replacement for packet transport; it enables learning traffic to become schedulable optical demand.

In an APN view, fronthaul, midhaul, and backhaul form a common optical substrate. For FL, the most immediate opportunity lies on the gNB-to-parameter-server path, where updates have already been aggregated into transfers with a known source, destination, approximate size, and deadline.

\begin{takeawaybox}
\textbf{Predictable Optical Transfer Requests.}
The real benefit of gNB aggregation is not simply fewer packets, but the creation of predictable, deadline-aware transfer requests. This enables federated learning traffic to be scheduled efficiently by an optical transport network.
\end{takeawaybox}

\begin{figure}[t]
    \centering
    \includegraphics[width=\linewidth]{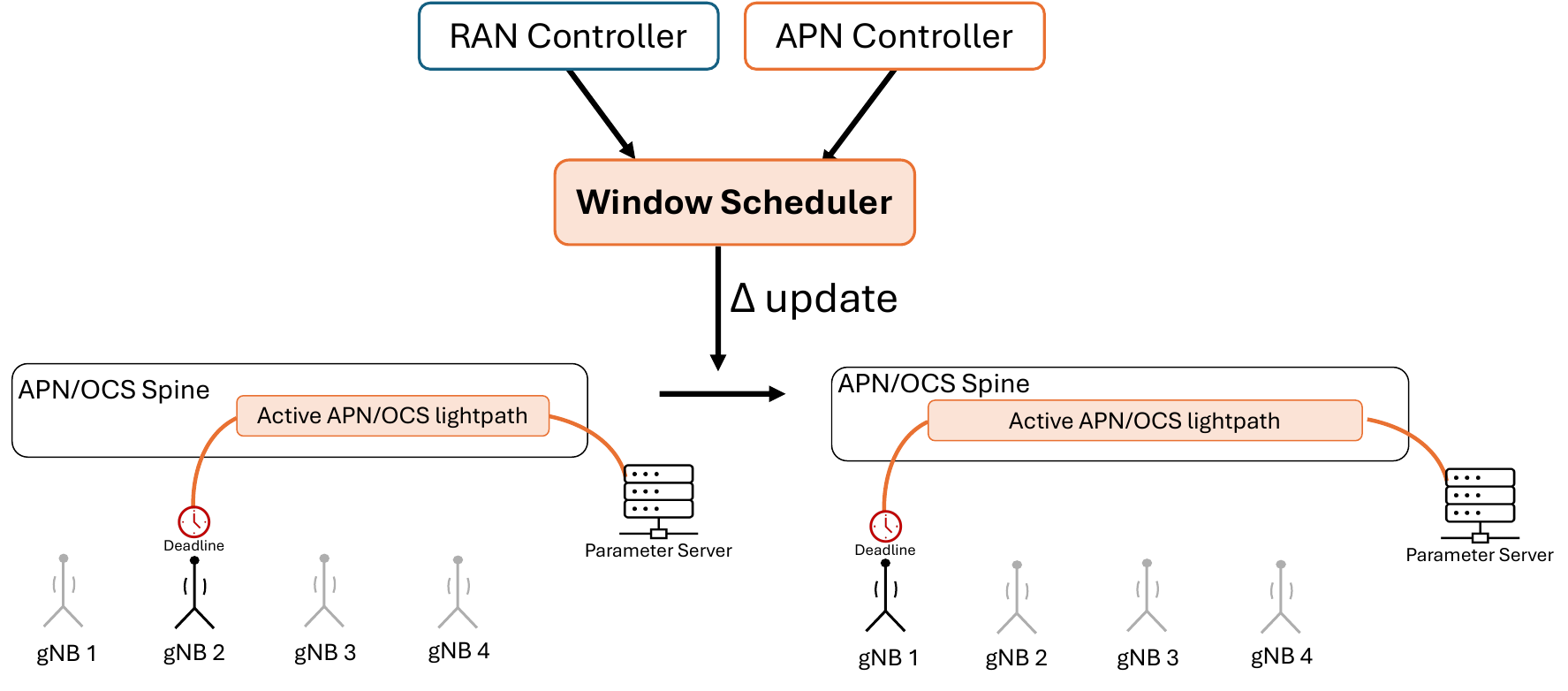}
    \caption{Learning-aware APN/OCS scheduling of gNB aggregate transfers.}
    \label{fig:learning-aware-apn-ocs}
\end{figure}

This enables a learning-aware control loop between the RAN and the optical layer. As illustrated in Fig.~\ref{fig:learning-aware-apn-ocs}, a Window Scheduler combines information from a RAN Controller, conceptually aligned with O-RAN RIC-based closed-loop control~\cite{polese2023understanding}, and an APN Controller that tracks optical resources~\cite{miyamoto2025progress}. Based on aggregate readiness, delivery deadlines, and resource availability, the scheduler assigns the active lightpath to the gNB whose transfer should be served next. In the example, the lightpath is initially assigned to gNB~2 and later reconfigured to gNB~1 as its aggregate approaches the deadline. This allows multiple gNBs to share a reconfigurable APN/OCS spine without dedicating a wavelength to each site. Rather than provisioning optical circuits for every UE update, the network schedules high-capacity circuits only for aggregated bursts after the gNB has absorbed device-level variability. The RAN therefore remains packet-switched for ordinary mobile services, while optical circuits are allocated dynamically to accelerate federated learning transfers.

\section{Learning and Transport Evaluation}
\label{sec:initial-results}

The previous sections introduced a learning-aware transport architecture that exploits the structure of federated learning traffic. In this section we evaluate its impact on learning performance, transport behavior, and optical provisioning. In particular, we use two complementary experiments to illustrate how gNB-side INA affects both federated learning and the traffic
offered to the transport network. The first experiment evaluates the
effect of INA on round duration and model convergence. The second
examines how INA reshapes asynchronous UE uploads into bounded
gNB-to-parameter-server transfers. Because the two experiments target
different aspects of the proposed architecture, they use different
numbers of UEs and model-update sizes.

\textbf{Learning-level impact.}
The first experiment uses the FLAG 5G user-plane emulator, which models
the SDAP, PDCP, and GTP-U layers together with MTU, queueing, and
wireless-loss effects~\cite{paolini2026programmable}. The scenario considers 60 UEs evenly distributed
across six gNBs. Each UE is assigned a 100-Mbps uplink, while the
gNB-to-parameter-server backhaul is constrained to 10~Mbps. The federated LLM fine-tuning workload generates LoRA adapter updates of approximately 6.28~MB. We define the \emph{round time} as the elapsed time from the beginning of local training until the participating UEs receive the global model for the following round. It therefore includes local training, update upload and aggregation, and global-model reception.

\begin{figure}[t]
    \centering
    \begin{minipage}[t]{0.49\linewidth}
        \centering
        \resizebox{\linewidth}{!}{%
            \input{figures/ue_accuracy_vs_time_cacm}
        }
        \vspace{0.3mm}
        {\scriptsize (a) Accuracy over elapsed time.}
    \end{minipage}%
    \hfill%
    \begin{minipage}[t]{0.49\linewidth}
        \centering
        \resizebox{\linewidth}{!}{%
            \input{figures/ue_time_decomposition_cacm}
        }
        \vspace{0.3mm}
        {\scriptsize (b) Round-time composition.}
    \end{minipage}
    \caption{Effect of gNB in-network aggregation on model accuracy and round duration.}
    \label{fig:flag-representative-results}
\end{figure}
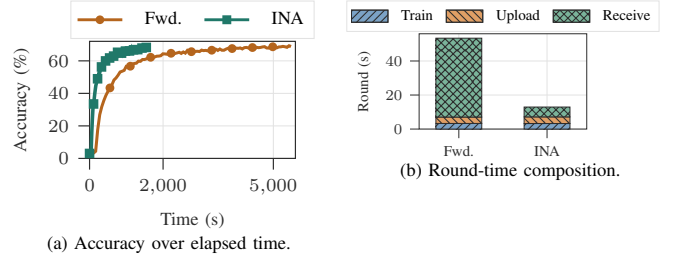

Figure~\ref{fig:flag-representative-results}(a) shows that INA
reaches the target model accuracy in less elapsed time than conventional
forwarding. Figure~\ref{fig:flag-representative-results}(b) explains
the source of this improvement. Local-training time remains comparable
because INA does not modify client computation. In contrast, the
communication component of the round is reduced because the gNB
combines UE updates before forwarding them over the constrained
backhaul. Shorter communication phases consequently produce shorter
rounds and faster model convergence over time.

\textbf{Transport-level impact.}
The same mechanisms also modify the traffic exposed to the backhaul.
FL flow identification distinguishes learning traffic from ordinary
user traffic, fragment-wise aggregation combines UE update streams,
and deadline closure bounds the release time of the resulting
aggregate. Together, these functions consolidate multiple asynchronous
UE-originated transfers into a single-source, single-destination
gNB-to-parameter-server demand with an estimable volume and a delivery
deadline inherited from the learning round. Without INA, each selected UE forwards its update independently.
The traffic observed on the backhaul is therefore arrival-driven and
depends on local-training completion times, radio scheduling,
retransmissions, and packetization. INA instead absorbs this
device-level variability at the gNB and releases one aggregate transfer
per learning round. The resulting traffic is not necessarily higher
rate than the superposition of UE flows, but it is more regular,
bounded, and temporally localized.

\begin{figure}[t]
    \centering
    \includegraphics[width=\linewidth]{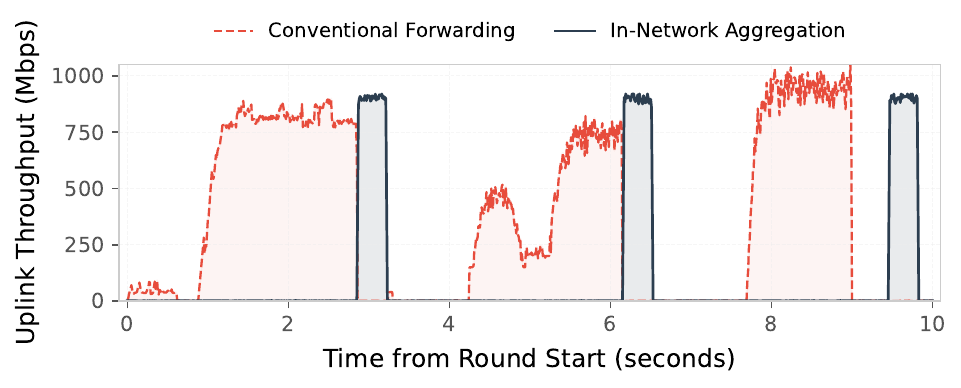}
    \caption{gNB-to-parameter-server traffic with conventional forwarding and in-network aggregation.}
    \label{fig:backhaul-traffic-comparison}
\end{figure}

Figure~\ref{fig:backhaul-traffic-comparison} reports the 
throughput observed on the gNB-to-parameter-server link over three
consecutive learning rounds. This experiment uses the OAI/FLAG
emulation environment with a cohort of 12 UEs, of which four are
selected in each round. Each selected UE transmits an update of
approximately 42.7~MB, representative of a LoRA PEFT adapter. With conventional forwarding, the backhaul carries four independent
updates per round, corresponding to approximately 170.8~MB of upstream
learning traffic. Because the updates become available at different
times, their transmissions extend over a relatively long and irregular
interval, while partial overlap among the flows produces throughput
fluctuations. With INA, the four updates are combined at the gNB and
only one aggregate, equal in size to a single 42.7-MB update, is
forwarded to the parameter server. INA therefore reduces the
gNB-to-parameter-server learning traffic by a factor of four. The aggregate appears as a shorter and more regular transfer approaching
1~Gbps, followed by a clearly identifiable idle interval. The transport
network consequently observes one bounded aggregate-transfer
opportunity per round rather than several asynchronous UE-originated
flows. INA therefore provides not only traffic reduction, but also the
timing information needed to coordinate transport provisioning.

\textbf{Deadline-aware optical provisioning.}
Provisioning an optical circuit is worthwhile only when its higher transmission capacity offsets the time required to establish or reconfigure the lightpath. Once a gNB closes its aggregate, the Window Scheduler knows its source, destination, size, readiness time, and latest useful delivery time. The APN Controller complements this information with the capacities of the available packet and optical paths and the optical setup time.

The following analysis assumes that INA has already combined the updates from each gNB into a single model-update-sized aggregate. The resulting gNB-level traffic can then be transmitted either over the existing packet path or a temporarily provisioned optical circuit, isolating the packet-versus-OCS transport decision. Let \(N_{\mathrm{gNB}}\) denote the number of active gNBs sharing the same transport segment. If each gNB generates one aggregate of size \(S\), the total traffic volume is

\begin{equation}
    B = N_{\mathrm{gNB}}S.
\end{equation}

The corresponding completion times are

\begin{equation}
    T_{\mathrm{pkt}} =
    \frac{8B}{C_{\mathrm{pkt}}},
    \qquad
    T_{\mathrm{ocs}} =
    T_{\mathrm{setup}} +
    \frac{8B}{C_{\mathrm{ocs}}},
\end{equation}

where \(C_{\mathrm{pkt}}\) and \(C_{\mathrm{ocs}}\) are the packet and optical capacities, and \(T_{\mathrm{setup}}\) is the optical setup time. OCS is beneficial whenever \(T_{\mathrm{ocs}} < T_{\mathrm{pkt}}\).

\begin{figure}[t]
    \centering
    \includegraphics[width=\linewidth]{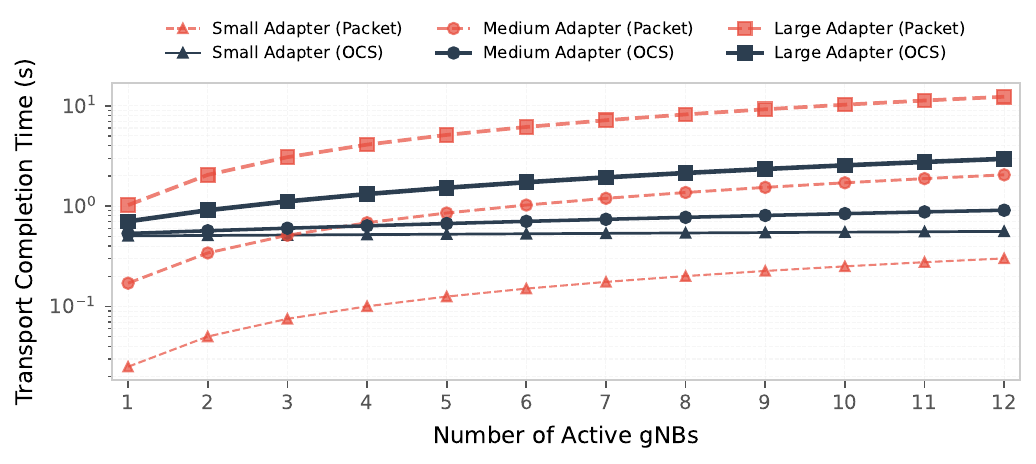}
    \caption{Packet-versus-OCS completion time as a function of the number of active gNBs for small, medium, and large PEFT adapter aggregates.}
    \label{fig:ocs-break-even}
\end{figure}

Figure~\ref{fig:ocs-break-even} illustrates this trade-off for PEFT adapters of 6.28 (Small), 42.7 (Medium), and 256~MB (Large), assuming 2-Gbps packet transport, 10-Gbps optical transport, and a 0.5-s setup time. Packet completion time increases linearly with traffic volume, whereas OCS incurs a fixed setup cost but benefits from higher capacity. Consequently, packet transport remains preferable for small adapters, OCS becomes advantageous from four active gNBs (170.8~MB) for medium adapters, and a single large aggregate already amortizes the setup delay.

More generally, optical provisioning should depend on the traffic available within a transmission window rather than on the number of active gNBs alone. Under the considered parameters, the packet--OCS break-even volume is approximately 156.25~MB, corresponding to one large aggregate, four medium aggregates, or about 25 small aggregates. This also highlights the role of the Window Scheduler: large aggregates can be transmitted immediately, whereas smaller ones may be grouped into a common transmission window whenever their delivery deadlines allow, amortizing optical setup costs without missing the learning deadline.

\section{Toward Deployable AI-Native RANs}
%\section{Challenges Ahead: From Shaped Traffic to Deployable Systems}

The proposed architecture illustrates how learning-aware aggregation can expose federated learning traffic as structured transport demand. Several research questions remain before this vision can become part of future AI-native RANs.

\textbf{RQ1: How should learning intent be exposed to the transport network?}
The optical layer does not need to understand the details of the learning algorithm, but it does require a compact description of each aggregate transfer. What information should the RAN expose? Is the source, destination, size, readiness time, and delivery deadline sufficient, or should future interfaces also communicate learning priorities, aggregation policies, or tolerance to delayed updates? Moreover, how should transport adapt when wireless variability prevents the planned transmission window from being met?

\textbf{RQ2: How can transport optimize learning without introducing bias?}
Communication decisions influence which clients ultimately contribute to the global model. Deadline-driven aggregation may consistently favor UEs with better radio conditions while excluding slower participants. How should transport scheduling account for fairness and statistical representativeness? Can mechanisms such as staleness-aware weighting, delayed aggregation, or partial contributions preserve model quality while still reducing communication latency?

\textbf{RQ3: When should optical transport be used?}
Packet transport remains the natural choice for control messages, model distribution, small updates, and fallback operation, whereas optical circuits become attractive only when sufficient traffic is available to amortize setup costs. How should a hybrid packet/optical infrastructure dynamically determine the operating point as workloads, network conditions, and optical resources evolve?

\textbf{RQ4: How should AI-native RANs be evaluated?}
The proposed architecture spans the learning system, the wireless network, and the transport infrastructure. Future evaluations should therefore move beyond isolated networking or machine learning metrics and instead consider the complete closed loop, including UE variability, gNB aggregation, transport scheduling, deadline satisfaction, communication efficiency, and model convergence.

\begin{takeawaybox}
{\bf  AI-Native RANs.} Ultimately, RAN Architectures should evolve from communication infrastructures into learning-aware systems that expose the semantics of distributed AI workloads to the transport network. Doing so enables networking decisions to optimize not only throughput and latency, but also the efficiency of the (distributed) learning process.
\end{takeawaybox}

\vspace{-3mm}
\section{Conclusion}
The main message of this article is not that optical circuits are always better than packet switching. 
Rather, APN/OCS becomes useful when the RAN first reshapes learning traffic into something that optical transport can schedule. 
This shifts the design question from ``how much capacity does the backhaul need?'' to ``which part of the learning process should the transport fabric see, and when?''
Future RANs that support distributed model adaptation will need to become more than passive carriers of model updates. 
They will need to act as learning-aware traffic shapers, exposing the right abstractions to packet and optical transport. 
The opportunity is to build a RAN in which computation, aggregation, and optical provisioning are coordinated parts of the same training system.
\bibliographystyle{IEEEtran}
\bibliography{references}

\vspace{-7mm}
\begin{IEEEbiographynophoto}
%[{\includegraphics[width=1in,height=1.25in,clip,keepaspectratio]{bio-pics/emilio.jpg}}]
{Emilio Paolini} received his Ph.D. degree in Emerging Digital Technologies, cum laude, from Scuola Superiore Sant'Anna, Pisa, Italy, in 2024. He is currently an Assistant Professor with the TeCIP Institute, Scuola Superiore Sant'Anna, Pisa, Italy. His research interests include AI-enabled NextG networks, programmable data planes, and federated learning. He received the GTTI Ph.D. Thesis Award for the Best Italian Ph.D. in Communication Technologies in 2024.
\end{IEEEbiographynophoto}
\vspace{-3mm}

\begin{IEEEbiographynophoto}%[{\includegraphics[width=1in,height=1.25in,clip,keepaspectratio]{bio-pics/emilio.jpg}}]\
{Andrea Pinto} received his BS and MS in computer engineering from the University of Naples Federico II. He graduated with a Ph.D. in Computer Science from Saint Louis University in 2026, specializing in computer networks, distributed systems, and scalable machine learning systems. His research focuses on distributed training and bandwidth-aware orchestration to alleviate network bottlenecks.
\end{IEEEbiographynophoto}

\vspace{-3mm}

\begin{IEEEbiographynophoto}%[{\includegraphics[width=1in,height=1.25in,clip,keepaspectratio]{bio-pics/emilio.jpg}}]
{Flavio Esposito} is an Associate Professor and Graduate Coordinator in the Department of Computer Science at Saint Louis University, where he also serves as a Fellow of the Research Institute. He received his Ph.D. in Computer Science from Boston University. His research focuses on edge computing, programmable and virtualized networks, network security, CPS, and applied artificial intelligence. % TOO LONG Dr. Esposito worked at Exegy, an high-frequency trading company in St.Louis, MO, and was a visiting researcher at Bell Laboratories, NJ, at Raytheon BBN Technologies, MA, at EURECOM, France and at the Center for Wireless Communications, Oulu, Finland. He is the recipient of several industry and NSF awards.
\end{IEEEbiographynophoto}

\begin{IEEEbiographynophoto}
%$$[{\includegraphics[width=1in,height=1.25in,clip,keepaspectratio]{bio-pics/emilio.jpg}}]
{Luca Valcarenghi} is a Full Professor at the Scuola Superiore Sant'Anna of Pisa, Italy, since 2024. He received the Ph.D. from UTD in 2001. Dr. Valcarenghi received a Fulbright Research Scholar Fellowship in 2009 and a JSPS "Invitation Fellowship Program for Research in Japan (Long Term)" in 2013. He coordinated, as a PI or local PI, several National and International projects. His main research interests are optical networks design, analysis, and optimization; energy efficiency in communications networks; optical access networks; zero touch network and service management; 5G technologies and beyond. 

\end{IEEEbiographynophoto}
\end{document}

%% file: figures/ue_accuracy_vs_time_cacm.tex
% CACM-style version derived from the original FLAG/TNSM data.
% Requires: \usepackage{pgfplots}
% Optional but recommended in preamble: \pgfplotsset{compat=1.18}

\begin{tikzpicture}[baseline=(current bounding box.north)]

\definecolor{cForward}{RGB}{181,101,29}
\definecolor{cINA}{RGB}{32,120,105}
\definecolor{cGrid}{RGB}{226,226,226}
\definecolor{cText}{RGB}{45,45,45}

\begin{axis}[
    width=\linewidth,
    height=3.15cm,
    xmin=0,
    xmax=5600,
    ymin=0,
    ymax=72,
    xlabel={Time (s)},
    ylabel={Accuracy (\%)},
    xtick={0,2000,5000},
    ytick={0,20,40,60},
    tick pos=left,
    tick align=outside,
    axis line style={draw=cText, line width=0.45pt},
    tick style={draw=cText, line width=0.45pt},
    tick label style={font=\scriptsize, text=cText},
    label style={font=\scriptsize, text=cText},
    grid=major,
    major grid style={draw=cGrid, line width=0.3pt},
    legend cell align={left},
    legend style={
        fill=white,
        fill opacity=1,
        draw=cGrid,
        text opacity=1,
        at={(0.5,1.02)},
        anchor=south,
        legend columns=2,
        font=\scriptsize,
        inner sep=2pt,
        column sep=6pt
    },
    legend image post style={line width=1.1pt},
    clip=false
]
\addplot[
    cForward,
    line width=1.15pt,
    mark=*,
    mark size=1.1pt,
    mark repeat=10,
    mark options={solid, fill=cForward, draw=cForward}
]
table {%
0 3.08393
55.1699534972509 3.11911166666667
110.499181115627 3.30373666666667
166.099520115058 4.40967833333333
221.091388837496 17.6378883333333
276.559203743935 26.9024766666667
332.311231267452 31.8213433333333
388.10216773351 35.5687483333333
443.929289182027 38.72102
499.246122193336 41.186255
554.470980274677 43.35372
609.548419825236 45.36171
664.572197743257 48.241005
720.018412339687 49.4720183333333
775.192261771361 51.1498783333333
830.556272939841 53.15548
885.886854565144 53.5967266666667
941.328878764312 54.1586766666667
997.194220725695 56.5855333333333
1052.83437974453 56.2617966666667
1108.48226967255 56.7098283333333
1164.02315391699 58.0407666666667
1219.11589208047 58.4020833333333
1274.50017738342 58.560345
1329.76868366003 59.8313366666667
1385.24334033728 59.7206216666667
1440.25179061095 60.83853
1495.58890163501 60.8193466666667
1551.06210952202 61.2729833333333
1606.23438140551 61.81455
1661.94173975388 62.200635
1717.50195809205 62.2589883333333
1772.68803532124 62.3793
1827.34996633132 62.8649083333333
1882.54562311967 62.9304583333333
1937.91129324834 63.155475
1994.31520169973 64.331735
2050.08390366236 64.2629966666667
2105.87747457027 64.0691433333333
2161.1737612923 64.2837783333333
2216.90031011899 64.667865
2272.34055645068 63.9616283333333
2328.05964312156 64.582335
2383.44245447715 64.89169
2438.6684820652 65.2705883333333
2494.18960835536 64.5639466666667
2549.87890062332 65.4512383333333
2604.63261189858 64.75779
2659.56186436017 66.4000783333333
2715.03264505068 65.3305416666667
2770.34089174668 65.643885
2825.44156701962 65.2881766666667
2880.78757047653 65.8149416666667
2936.02496128082 65.888885
2991.79719730218 65.717425
3047.52341815233 66.6119133333333
3102.76336398919 66.29497
3157.86788106362 66.447645
3213.74501569271 66.1594683333333
3268.61247463624 66.537175
3323.90955786308 66.54956
3378.81615207593 66.483615
3433.89299009244 67.1830516666667
3489.65980784496 66.4064766666667
3544.88365234534 67.2845683333333
3600.1391221563 66.53637
3655.32956331968 66.5295783333333
3710.75710472663 67.21263
3765.79963281552 67.7278183333333
3821.09252824783 66.9652283333333
3877.43457265298 67.5639533333333
3932.76195918322 67.54557
3987.87520682812 67.840525
4042.86133236885 67.7190266666667
4098.61195148627 67.9508383333333
4154.80432113806 67.3353316666667
4210.0166973273 68.1143066666667
4265.03360897303 68.2669866666667
4320.26950005293 67.96323
4375.99433891376 67.2953616666667
4431.10906658173 68.1942433333333
4486.29612737099 67.2733833333333
4541.45082749923 68.304555
4596.66483054956 68.116705
4652.21266988516 68.2581933333333
4707.13069788615 68.5703366666667
4762.06359538237 68.7314183333333
4817.57348622878 67.9280583333333
4872.64261159897 67.8493216666667
4927.81994743745 68.8137533333333
4983.52264428139 68.6542716666667
5038.80182071527 68.58154
5093.59775282939 68.02438
5148.67179139455 68.559155
5203.93538109461 68.6279016666667
5259.20747790734 68.906875
5314.56847137213 68.669065
5370.54193421205 68.317745
5426.18072435856 69.29496
5481.23535498778 68.8585166666667
};
\addlegendentry{Fwd.}

\addplot[
    cINA,
    line width=1.35pt,
    mark=square*,
    mark size=1.15pt,
    mark repeat=7,
    mark options={solid, fill=cINA, draw=cINA}
]
table {%
0 3.04556354916067
15.609395925204 3.10191846522782
31.0748208284378 3.64828137490008
46.1591405471166 6.37290167865707
61.6831295092901 16.8493205435651
77.5507851759593 24.6298960831335
93.0617445707321 30.9580335731415
108.728647645315 33.4828137490008
123.875729719798 36.0623501199041
139.427148354053 39.4060751398881
155.182975172997 42.3253397282174
170.61354812781 44.8033573141487
186.049400913715 46.1714628297362
202.234143459797 48.4936051159073
217.808999200662 48.982414068745
233.312301003933 50.3365307753797
248.4153835694 51.6742605915268
263.559412395954 51.3756994404476
278.667495961984 52.9704236610711
294.201763923963 54.6290967226219
309.441199469566 55.4052757793765
324.686304950714 56.2593924860112
343.492409658432 57.636690647482
359.257268369198 57.3185451638689
374.355621735255 59.2561950439648
389.492114051183 58.8369304556355
404.625863011678 59.8976818545164
420.12470053037 60.0395683453237
435.698310875893 59.96482813749
451.097129555543 61.4044764188649
466.526599685351 60.7494004796163
482.118003773689 60.8157474020783
497.29946971337 61.2370103916867
512.63976517121 61.6227018385292
528.183117198944 62.0991207034373
545.713033715884 61.777378097522
561.007160600026 61.6111111111111
576.202598865827 62.1870503597122
592.236198496819 61.621103117506
607.907298183441 63.0603517186251
622.926230990887 62.8101518784972
638.222807395458 63.3369304556355
654.040293979645 62.9272581934452
669.53543252945 64.1618705035971
684.387276506424 63.8313349320544
699.725154248873 63.5959232613909
715.566511436303 63.6514788169464
730.897991077105 63.7725819344524
746.465663878123 64.5931254996003
762.221974817912 65.2721822541966
777.724966402849 64.3996802557954
793.196931167444 65.3565147881695
808.293757549922 65.3840927258193
823.584662501017 64.6007194244604
839.069279456139 64.6590727418066
854.276026574771 65.5243804956035
869.61494636933 64.7378097521982
884.686642162005 64.1426858513189
900.151686914762 65.1546762589928
916.357507582505 65.7769784172662
931.998448499044 65.8173461231015
946.773020410538 65.9764188649081
961.885245732466 65.2625899280576
977.442380841573 66.019584332534
994.784385049343 65.3852917665867
1009.94232252439 66.4216626698641
1025.50105532805 65.8856914468425
1040.81098616521 66.3836930455635
1056.04954727491 66.5579536370903
1071.26123148203 66.7989608313349
1086.36674877405 66.0943245403677
1101.93558034102 66.2801758593125
1117.84424695969 67.1115107913669
1133.76183005174 66.7769784172662
1149.46449852387 67.7222222222222
1165.2097802043 66.8677058353317
1181.27645408312 66.9464428457234
1196.82497713168 66.7797761790568
1212.8994283994 67.3137490007994
1229.28189157248 67.2078337330136
1245.00954029957 67.171862509992
1261.37805054188 67.3745003996803
1277.77348656654 66.8001598721023
1293.96372823318 66.9212629896083
1310.25997674068 66.9072741806555
1326.89570978085 67.3213429256595
1343.62960613966 66.9796163069544
1360.14388465484 66.6614708233413
1377.05291078885 67.6254996003197
1393.37099664211 67.7513988808953
1409.85998452902 67.4136690647482
1426.16917476257 67.8772981614708
1442.6575323383 67.3077537969624
1459.54700802962 66.9384492406075
1475.81284976403 67.4104716227018
1492.47678909302 67.9248601119105
1509.60850867033 67.8892885691447
1526.27512964805 67.6706634692246
1543.18546503385 68.2633892885691
1560.65552625656 67.8193445243805
};
\addlegendentry{INA}

\end{axis}
\end{tikzpicture}

%% file: figures/ue_time_decomposition_cacm.tex
% CACM-style version derived from the original FLAG/TNSM data.
% Requires: \usepackage{pgfplots}
% Requires for hatching: \usetikzlibrary{patterns}

\begin{tikzpicture}[baseline=(current bounding box.north)]

\definecolor{cTrain}{RGB}{86,136,180}
\definecolor{cUpload}{RGB}{215,132,58}
\definecolor{cReceive}{RGB}{70,150,115}
\definecolor{cGrid}{RGB}{226,226,226}
\definecolor{cText}{RGB}{45,45,45}

\begin{axis}[
    width=\linewidth,
    height=3.15cm,
    xmin=-0.45,
    xmax=1.45,
    ymin=0,
    ymax=56,
    ylabel={Round (s)},
    xtick={0,1},
    xticklabels={Fwd.,INA},
    ytick={0,20,40},
    tick pos=left,
    tick align=outside,
    axis line style={draw=cText, line width=0.45pt},
    tick style={draw=cText, line width=0.45pt},
    tick label style={font=\scriptsize, text=cText},
    label style={font=\scriptsize, text=cText},
    grid=major,
    major grid style={draw=cGrid, line width=0.3pt},
    legend cell align={left},
    legend style={
        fill=white,
        fill opacity=1,
        draw=cGrid,
        text opacity=1,
        at={(0.5,1.02)},
        anchor=south,
        legend columns=3,
        font=\scriptsize,
        inner sep=2pt,
        column sep=4pt
    },
    clip=false
]

% Forwarding values: training=3.3057, upload=3.7606, receive=46.3346
% INA values:        training=3.2530, upload=3.9017, receive=5.7889

% Training
\draw[draw=cText, line width=0.35pt, fill=cTrain!78,
      postaction={pattern=north east lines, pattern color=cText}]
    (axis cs:-0.26,0) rectangle (axis cs:0.26,3.30572768791517);
\draw[draw=cText, line width=0.35pt, fill=cTrain!78,
      postaction={pattern=north east lines, pattern color=cText}]
    (axis cs:0.74,0) rectangle (axis cs:1.26,3.25299534664118);
\addlegendimage{area legend, draw=cText, fill=cTrain!78,
      postaction={pattern=north east lines, pattern color=cText}}
\addlegendentry{Train}

% Upload
\draw[draw=cText, line width=0.35pt, fill=cUpload!78,
      postaction={pattern=north west lines, pattern color=cText}]
    (axis cs:-0.26,3.30572768791517) rectangle (axis cs:0.26,7.0663113070329);
\draw[draw=cText, line width=0.35pt, fill=cUpload!78,
      postaction={pattern=north west lines, pattern color=cText}]
    (axis cs:0.74,3.25299534664118) rectangle (axis cs:1.26,7.15464561019058);
\addlegendimage{area legend, draw=cText, fill=cUpload!78,
      postaction={pattern=north west lines, pattern color=cText}}
\addlegendentry{Upload}

% Receive / downstream wait
\draw[draw=cText, line width=0.35pt, fill=cReceive!72,
      postaction={pattern=crosshatch, pattern color=cText}]
    (axis cs:-0.26,7.0663113070329) rectangle (axis cs:0.26,53.4009459828536);
\draw[draw=cText, line width=0.35pt, fill=cReceive!72,
      postaction={pattern=crosshatch, pattern color=cText}]
    (axis cs:0.74,7.15464561019058) rectangle (axis cs:1.26,12.9435393488053);
\addlegendimage{area legend, draw=cText, fill=cReceive!72,
      postaction={pattern=crosshatch, pattern color=cText}}
\addlegendentry{Receive}

\end{axis}
\end{tikzpicture}